\documentclass{bmvc2k}
\usepackage{enumitem}
\usepackage{amsfonts}
\usepackage{multirow}
\usepackage{float}
\usepackage{wrapfig}

\usepackage{amsmath,amsfonts,bm}

\renewcommand\cdots{...}

\newcommand{\vx}{\mathbf{x}}

\newcommand{\mbr}[1]{\mathbb{R}^{#1}}

\newcommand{\vz}{\mathbf{z}}

\newcommand{\vw}{\boldsymbol{w}}

\newcommand{\stkout}[1]{{\ifmmode\text{\sout{\ensuremath{#1}}}\else\sout{#1}\fi}}

\title{FreqSelect: Frequency-Aware fMRI-to-Image Reconstruction}

\addauthor{Junliang Ye}{u7727288@anu.edu.au}{1}
\addauthor{Lei Wang}{l.wang4@griffith.edu.au}{\textsuperscript{$\dag$}, 2, 3}
\addauthor{Md Zakir Hossain}{zakir.hossain1@curtin.edu.au}{1, 3, 4}

\addinstitution{
 Australian National University
}
\addinstitution{
 Griffith University
}
\addinstitution{
 Data61/CSIRO
}
\addinstitution{
 Curtin University
}

\runninghead{Ye, Wang, Hossain}{Frequency-Aware fMRI-to-Image Reconstruction}

\def\eg{\emph{e.g}\bmvaOneDot}

\def\etal{\emph{et al}\bmvaOneDot}

\begin{document}

\renewcommand{\thefootnote}{\fnsymbol{footnote}}
\footnotetext[2]{Corresponding author.}
\renewcommand{\thefootnote}{\arabic{footnote}}

\maketitle

\begin{abstract}
Reconstructing natural images from functional magnetic resonance imaging (fMRI) data remains a core challenge in natural decoding due to the mismatch between the richness of visual stimuli and the noisy, low resolution nature of fMRI signals.
While recent two-stage models, combining deep variational autoencoders (VAEs) with diffusion models, have advanced this task, they treat all spatial-frequency components of the input equally.
This uniform treatment forces the model to extract meaning features and suppress irrelevant noise simultaneously, limiting its effectiveness.
We introduce \textit{FreqSelect}, a lightweight, adaptive module that selectively filters spatial-frequency bands before encoding.
By dynamically emphasizing frequencies that are most predictive of brain activity and suppressing those that are uninformative, FreqSelect acts as a content-aware gate between image features and natural data. It integrates seamlessly into standard very deep VAE-diffusion pipelines and requires no additional supervision.
Evaluated on the Natural Scenes dataset, FreqSelect consistently improves reconstruction quality across both low- and high-level metrics.
Beyond performance gains, the learned frequency-selection patterns offer interpretable insights into how different visual frequencies are represented in the brain.
Our method generalizes across subjects and scenes, and holds promise for extension to other neuroimaging modalities, offering a principled approach to enhancing both decoding accuracy and neuroscientific interpretability.

\end{abstract}
\section{Introduction}
Decoding visual experiences from functional magnetic resonance imaging (fMRI) is a central challenge at the intersection of neuroscience and machine learning. Natural images span a wide spectrum of spatial frequencies, from coarse structures to fine-grained textures, whereas fMRI signals are inherently noisy, temporally delayed, and spatially blurred due to the low resolution and hemodynamic nature of the measurements.
This mismatch creates a fundamental obstacle: how can we recover the full richness of visual perception from signals that are both incomplete and noisy?

%
Recent advances have made notable progress by using two-stage generative pipelines.
These approaches typically use a very deep variational autoencoder (VDVAE)~\cite{child2021vdvae} to produce a coarse reconstruction and then refine the semantic details using a latent diffusion model.
Methods like Brain-Diffuser~\cite{ozcelik2023naturalscene} and MindDiffuser~\cite{lu2023minddiffuser} have shown impressive perceptual quality in image reconstruction tasks.
However, they treat all spatial-frequency components of the visual input equally. This uniform treatment forces the model to simultaneously suppress noise and extract meaningful features across the entire frequency spectrum, leading to suboptimal use of model capacity and potential degradation in reconstruction fidelity.



In this work, we propose \textit{FreqSelect}, a lightweight and adaptive frequency-selection module that filters spatial-frequency components of the input image prior to encoding. Rather than applying a fixed or uniform filter, FreqSelect dynamically adjusts the emphasis placed on different frequency bands based on their relevance to the underlying brain activity. 
It acts as a content-aware gate, suppressing uninformative frequencies while preserving those most predictive of neural responses. Crucially, FreqSelect integrates seamlessly into existing VDVAE–diffusion pipelines and introduces no need for additional supervision.

We evaluate FreqSelect on the Natural Scenes Dataset and show that it not only enhances reconstruction quality across both low- and high-level metrics, but also offers new insights into how different spatial frequencies are represented in the human brain.
Our main contributions are:
\renewcommand{\labelenumi}{\roman{enumi}.}
\begin{enumerate}[leftmargin=0.6cm]
\item \textbf{Adaptive frequency selection.} We introduce a dynamic, data-driven module that learns to gate spatial-frequency bands based on their informativeness for fMRI decoding, enabling better alignment between visual inputs and neural signals.

\item \textbf{Improved reconstruction quality.} FreqSelect leads to consistent performance improvements in image reconstruction tasks by reducing the burden on downstream models to separate noise from signal.

\item \textbf{Neuroscientific insight.} The learned frequency‐selection patterns provide interpretable evidence about how the brain represents visual information at different spatial scales, opening new avenues for neuroscientific discovery.
\end{enumerate}

In Appendix~\ref{app:relatedwork}, we review closely related work and highlight the key distinctions between our approach and prior methods. Below, we introduce our proposed method.

\begin{figure}[tbp]
\centering
\begin{tabular}[t]{cc}
\subfigure[Image and its frequency spectrum.]{\label{fig:spatial-frequency}\includegraphics[width=0.45\textwidth]{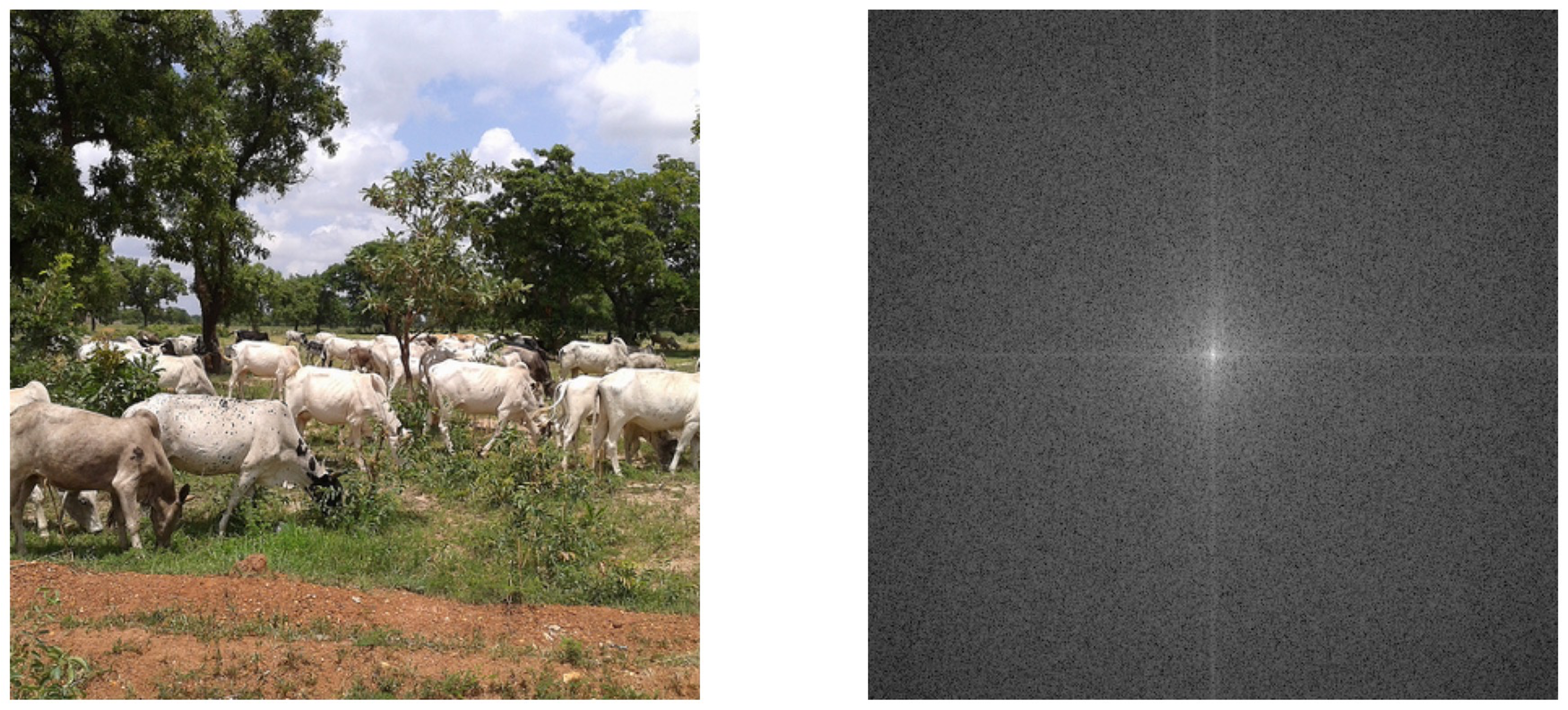}} &
\subfigure[Band-pass mask and filtered output.]{\label{fig:band-pass-example}\includegraphics[width=0.45\textwidth]{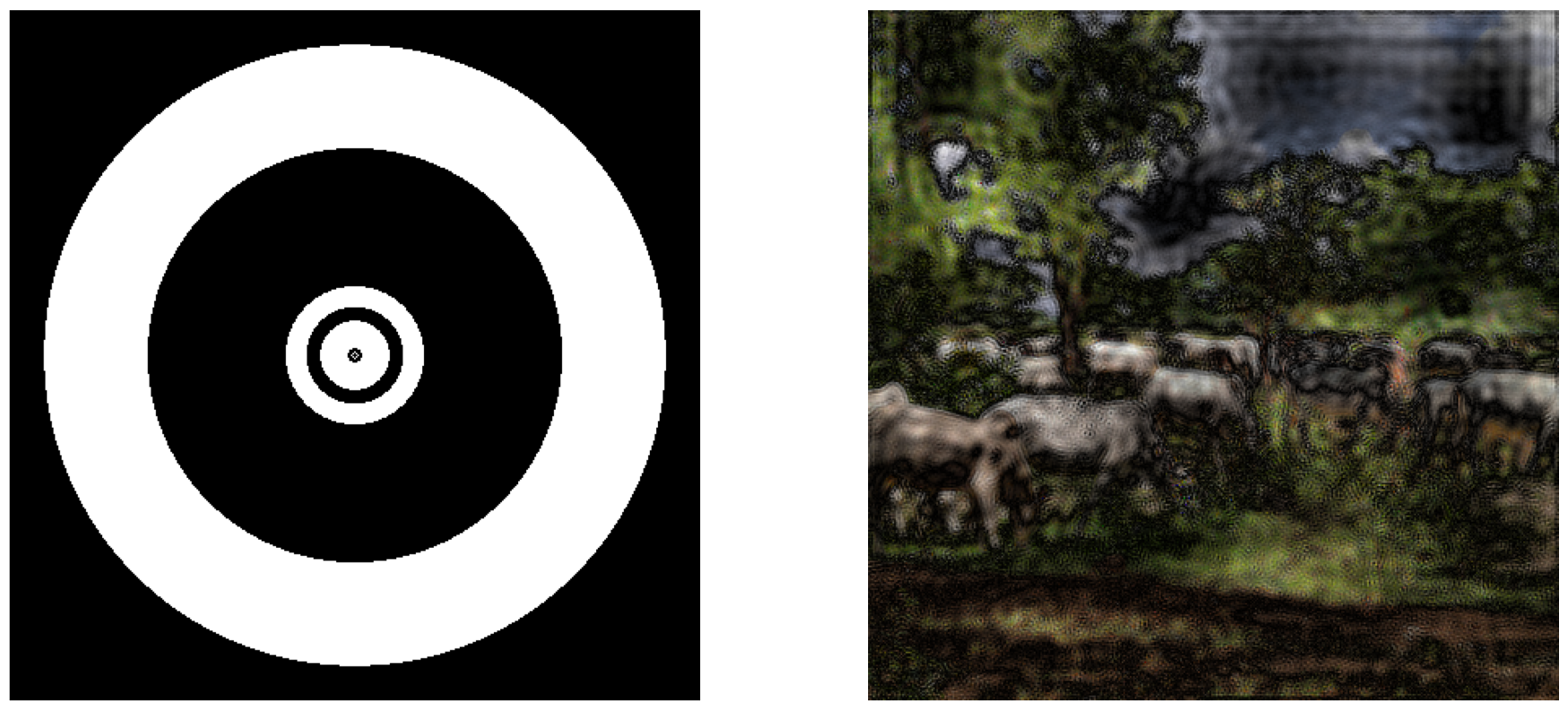}}
\end{tabular}
\caption{(a) A sample image and its frequency-domain representation, where low-frequency components cluster near the center and high-frequency details appear toward the edges. (b) A circular band-pass mask highlighting intermediate frequencies (white = passed, black = suppressed), and the resulting spatial-domain reconstruction via inverse FFT, which preserves mid-scale structures while attenuating low-frequency shapes and high-frequency noise.}
\label{fig:loss-comp}
\end{figure}

\begin{figure}[tbp]
    \centering
    \includegraphics[width=0.95\linewidth]{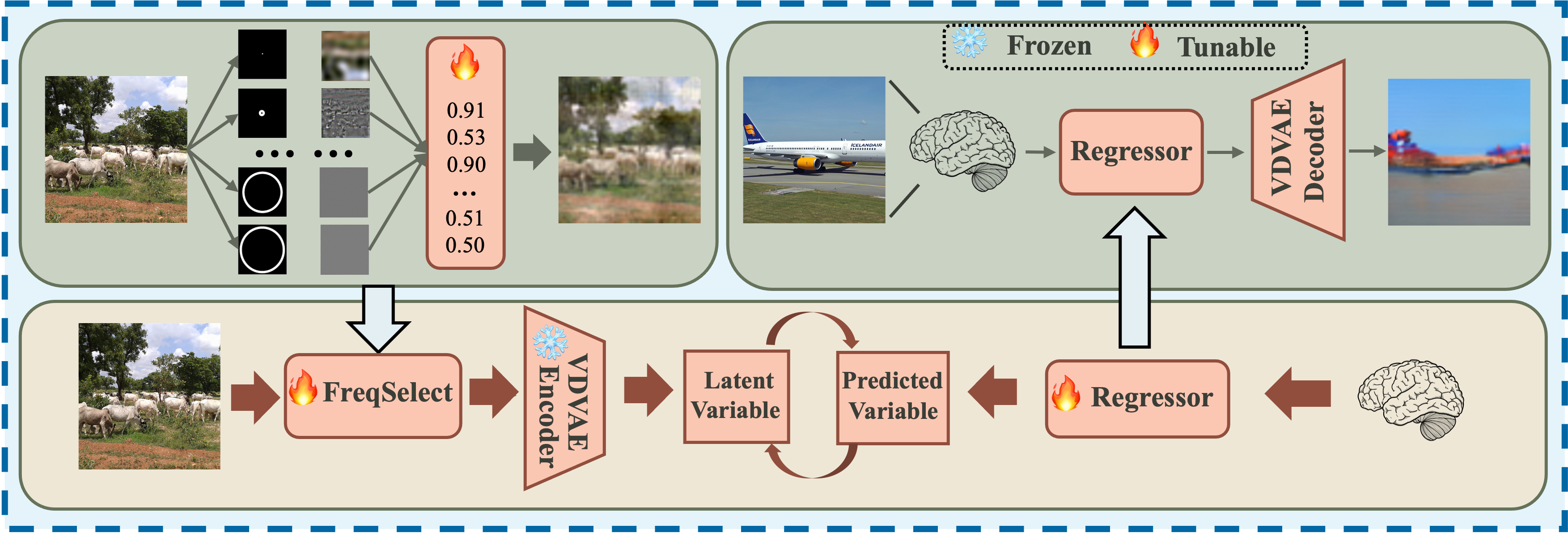}
    \caption{
Overview of our fMRI‐to‐image reconstruction pipeline with FreqSelect (Stage 1). 
During training, the input image is first processed by \textit{FreqSelect} (top left): An input image is decomposed into $N$ frequency bands via circular masks in the Fourier domain. Each band is transformed back to the spatial domain, weighted, and summed to form a filtered image retaining informative frequencies.
The resulting frequency-filtered image is then passed through a frozen  very deep variational autoencode (VDVAE) encoder to obtain the ``true'' latent representation $\vz_{\rm true}$.
In parallel, a ridge regressor maps fMRI signals to a predicted latent vector $\vz_{\rm pred}$, and the model is optimized by minimizing the latent-space loss $\|\vz_{\rm true} - \vz_{\rm pred}\|^2$.
At inference time, the trained regressor (top right) generates $\vz_{\rm pred}$ directly from fMRI data, which is decoded by the frozen VDVAE decoder to reconstruct a low-level image. This output then serves as the input to Stage 2 of the pipeline.
}
    \label{fig:stage1}
\end{figure}



\section{Methodology}

In this section, we present our complete fMRI-to-image reconstruction pipeline, with particular emphasis on the role and integration of the proposed \textit{FreqSelect} module.

\subsection{FreqSelect: Frequency-Aware Adaptive Filtering}

\textbf{Input and frequency transform.}  
Given an input image $\vx \in \mathbb{R}^{C \times H \times W}$, where $C = 3$ denotes the RGB color channels and $H = W = 64$ is the image resolution, we begin by transforming each channel into the frequency domain via the 2D Discrete Fourier Transform (DFT):
\begin{equation}
\widehat{\vx}_c(u,v) = \sum_{n=0}^{H-1} \sum_{m=0}^{W-1} \vx_c(n,m)\, e^{-j2\pi\left(\frac{u n}{H} + \frac{v m}{W}\right)},
\end{equation}
for each channel $c \in \{1, 2, 3\}$. To facilitate radial filtering, we apply an \texttt{fftshift} operation to center zero-frequency component. The radial distance from the center is then given by:
\begin{equation}
r = \sqrt{\left(u - \tfrac{H}{2}\right)^2 + \left(v - \tfrac{W}{2}\right)^2},
\end{equation}
so that small $r$ values correspond to low frequencies and large $r$ values to high frequencies. Figure~\ref{fig:spatial-frequency} visualizes this transformation.

\textbf{Band-pass decomposition.} To isolate distinct frequency components, we divide the frequency space into $N$ radial bands, using cutoff values $\{\nu_i\}_{i=0}^N$ with $\nu_0 = 0$ and $\nu_N$ equal to the Nyquist frequency. These values are linearly spaced across the radial frequency range. For each band $i$, we define a binary mask:
\begin{equation}
M_i(u,v) = 
\begin{cases}
1, & \nu_{i-1} < r \le \nu_i, \\
0, & \text{otherwise}.
\end{cases}
\end{equation}

This mask selects a ring-shaped frequency band. Applying it to the DFT of the image and then performing an inverse DFT yields the corresponding band-limited image:
\begin{equation}
f_i(x) = \mathrm{IDFT}\left(M_i \cdot \widehat{\vx}\right).
\end{equation}
This process produces a series of filtered images, each emphasizing specific frequency bands. Figure~\ref{fig:band-pass-example} illustrates an example mask and its effect on spatial content.

\textbf{Adaptive frequency weighting and fusion.} To prioritize the most informative frequency bands, we introduce a learnable scalar weight $w_i$ for each band. These are passed through a sigmoid to ensure values between 0 and 1: $\alpha_i = \sigma(w_i) \in (0, 1)$.
The final filtered image is then computed as a weighted average of all band-pass outputs:
\begin{equation}
\widetilde{\vx} = \frac{\sum_{i=1}^N \alpha_i \, f_i(\vx)}{\sum_{i=1}^N \alpha_i + \varepsilon},
\end{equation}
where $\varepsilon$ is a small constant (\eg, $10^{-10}$) to prevent division by zero. 
This mechanism enables the model to adaptively emphasize frequency bands most predictive of neural responses. Figure~\ref{fig:stage1} (top left) shows the entire FreqSelect pipeline.

\subsection{Low-Level Image Reconstruction Using FreqSelect} 

Variational Autoencoders (VAEs) model complex data distributions by encoding an input $\vx$ into a low-dimensional latent representation $\vz$, typically under a Gaussian prior, and then decoding it to reconstruct $\vx$. 
While conventional VAEs often fail to capture the full diversity and structure of natural scenes, VDVAE~\cite{child2021vdvae} addresses this limitation through a hierarchy of latent variables that progressively model finer-grained spatial details.
Specifically, VDVAE defines a structured approximate posterior:
\begin{equation}
    q_\phi(\vz \mid \vx)
= q_\phi(\vz_0 \mid \vx)\,
  q_\phi(\vz_1 \mid \vz_0, \vx)\,\cdots\,
  q_\phi(\vz_K \mid \vz_{K-1}, \vx),
\end{equation}
and a corresponding top-down generative prior:
\begin{equation}
    p_\theta(\vz)
= p_\theta(\vz_0)\,
  p_\theta(\vz_1 \mid \vz_0)\,\cdots\,
  p_\theta(\vz_K \mid \vz_{K-1}).
\end{equation}

Here, $\vz_0$ represents a coarse, low-resolution latent code, while deeper layers $\vz_i$ ($i > 0$) refine the reconstruction by encoding increasingly fine-grained features. This hierarchical structure allows the model to generate rich, high-fidelity representations of natural images.

To maintain consistency with prior work, we adopt the exact VDVAE configuration used in brain-diffuser~\cite{ozcelik2023naturalscene}. Specifically, we use a pretrained VDVAE trained on $64 \times 64$ ImageNet images, comprising 75 latent layers. However, following~\cite{ozcelik2023naturalscene}, we extract latents from only the first 31 layers, as deeper layers were found to contribute minimal additional benefit. These 31 latent outputs are concatenated into a single 91,168-dimensional feature vector.

During training, each stimulus image is passed through the frozen VDVAE encoder to obtain its latent representation. A ridge regression model is then trained to predict this latent vector from the corresponding fMRI pattern. At inference time, the trained regressor generates a latent vector from a novel fMRI input, which is decoded by the frozen VDVAE decoder to produce a $64 \times 64$ image reconstruction. This coarse reconstruction serves as the initial input to the second-stage refinement module, based on latent diffusion.

\textbf{Integration of FreqSelect.} Figure~\ref{fig:stage1} provides an overview of  incorporating FreqSelect. 
The fused image $\widetilde{\vx}$ is permuted and normalized to match the input format of the pretrained VDVAE encoder. This encoder produces a hierarchical set of latent variables $\{\vz_k\}_{k=0}^{K-1}$, which are flattened and concatenated into a single latent vector: $\vz_{\mathrm{true}} \in \mathbb{R}^D$.
This representation serves as the supervision target for learning to decode brain activity.

\textbf{Stage 1 training objective.}  
We train \textit{FreqSelect} and the fMRI-to-latent regressor jointly, minimizing the latent space mean squared error:
\begin{equation}
\mathcal{L}_{\mathrm{s1}} = \frac{1}{B} \sum_{b=1}^B \left\| \vz_{\mathrm{true}}^{(b)} - \vz_{\mathrm{pred}}^{(b)} \right\|^2,
\end{equation}
where $B$ is the batch size. 
During training, the VDVAE encoder is frozen. Only the frequency weights $\vw \in \mbr{N}$ and regressor parameters are updated. This setup encourages the model to learn which frequency bands are most informative for fMRI decoding.

\textbf{Stage 1 inference procedure.}  
At test time, unseen fMRI samples are mapped through the trained regressor to predict latent codes $\vz_{\mathrm{pred}}$, which are then decoded by the frozen VDVAE decoder to reconstruct full images. This enables robust, frequency-aware brain-to-image decoding that uses the structure of the visual frequency spectrum.

\subsection{High-Level Refinement via Latent Diffusion}

\begin{wrapfigure}{r}{0.5\linewidth}
    \centering
    \includegraphics[width=\linewidth]{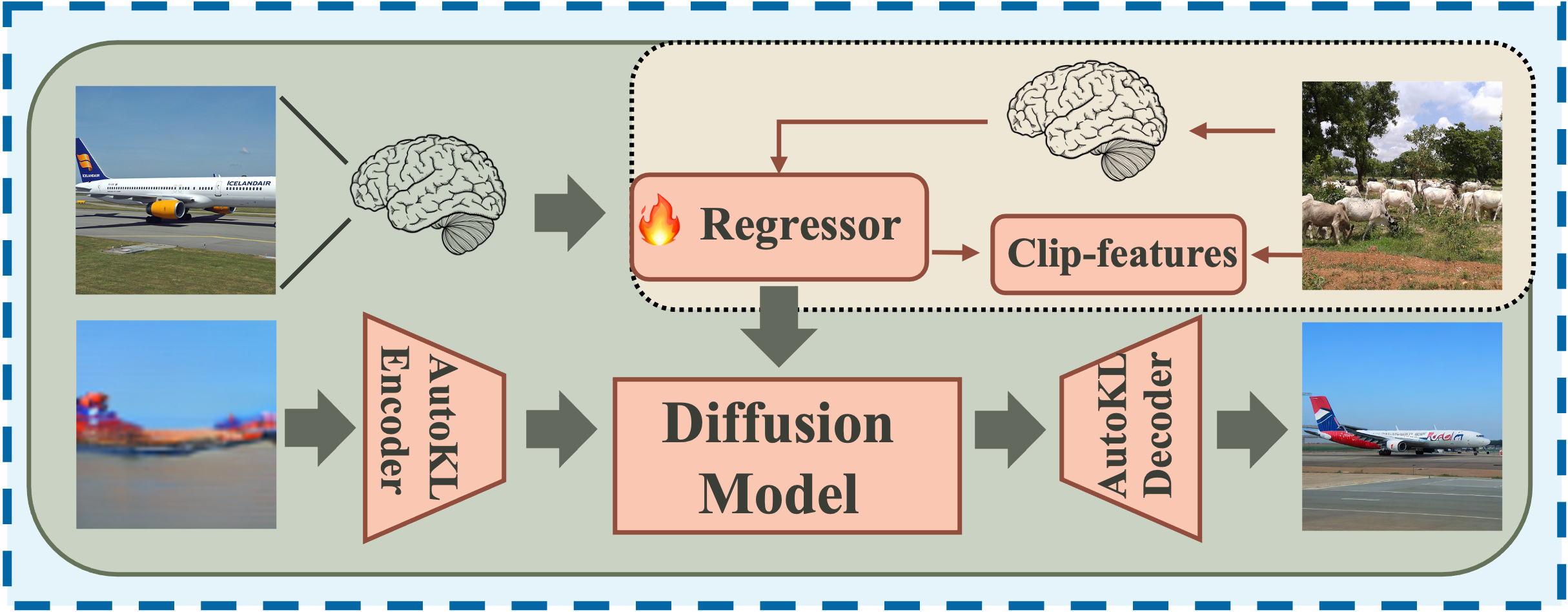}
    \caption{Stage 2: A latent diffusion model refines the VDVAE reconstruction using predicted CLIP features from fMRI, injecting high-level semantics and structure, for high-level image reconstruction.}
    \label{fig:stage2}
\end{wrapfigure}

While the VDVAE encoder produces a coherent low-level layout from fMRI, it is limited in capturing high-level semantic content and photorealistic textures. 
To address this shortcoming and ensure consistency with baseline methods, we adopt the Versatile Diffusion latent diffusion model~\cite{xu2023versatile} as a second-stage refinement module (see Fig. \ref{fig:stage2}). 
This model is pretrained on the LAION-2B-en dataset~\cite{schuhmann2022laion5b} at $512 \times 512$ resolution, using CLIP-ViT/L-14~\cite{radford2021learning} to separately extract image and text embeddings.

\textbf{Stage 2 training objective.} During training, we freeze all components of the diffusion model, including U-Net~\cite{ronneberger2015u} and AutoKL~\cite{Rombach_2022_CVPR} modules. The training objective is to minimize standard noise-prediction loss:
\begin{equation}
    \mathcal{L}_{\mathrm{s2}} = \mathbb{E}_{t, \vz_0, \epsilon, y} \left\| \epsilon - \epsilon_\theta(\vz_t, t, \tau_\phi(y)) \right\|^2,
\end{equation}
where $\epsilon$ is used as ground truth noise that is added to $\vz_0$, $\epsilon_\theta(\cdot)$ is the model's prediction of noise. This loss compares the true injected noise $\epsilon$ to the predicted noise $\epsilon_\theta$, training the model to denoise $\vz_t$ back toward $\vz_0$. 
The noisy latent at timestep $t$ is generated as:
\begin{equation}
    \vz_t = \sqrt{\bar\beta_t} \, \vz_0 + \sqrt{1 - \bar\beta_t} \, \epsilon, 
\end{equation}

Here, $\vz_0$ is the initial latent representation obtained from AutoKL encoder, $\epsilon$ is a sample from a standard normal distribution: $\mathcal{N}(0, I)$, and $\tau_\phi(y)$ represents the CLIP-based conditioning features predicted from fMRI using a trained ridge regressor. $\bar\beta_t$ denotes the cumulative product of noise scaling factors up to timestep $t$, defined as $\bar\beta_t = \prod_{s=1}^{t} \beta_s$, and reflects the total signal retention after $t$ diffusion steps.
These embeddings can originate from either the target image or its caption, enabling semantic supervision from multiple modalities.


\textbf{Stage 2 inference procedure} At inference, the $64 \times 64$ VDVAE output from Stage 1 is upsampled to $512 \times 512$ and encoded by the frozen AutoKL encoder to obtain $\vz_0$. We apply forward diffusion for $T_{\mathrm{init}} = 37$ steps ($\sim$75\% of the 50-step schedule), yielding a noisy latent $\vz_{T_{\mathrm{init}}}$. Reverse diffusion is then performed with cross-attention to predicted CLIP embeddings, and the denoised latent is decoded by AutoKL to produce a high-fidelity $512 \times 512$ reconstruction enriched with semantic detail, color, and structure, significantly surpassing VDVAE alone. 
We present our experiments and evaluations below.

\begin{figure}[tbp]
\centering
\begin{tabular}[t]{cc}
\subfigure[]{\label{fig:Training-Time Dynamics}\includegraphics[width=0.45\textwidth]{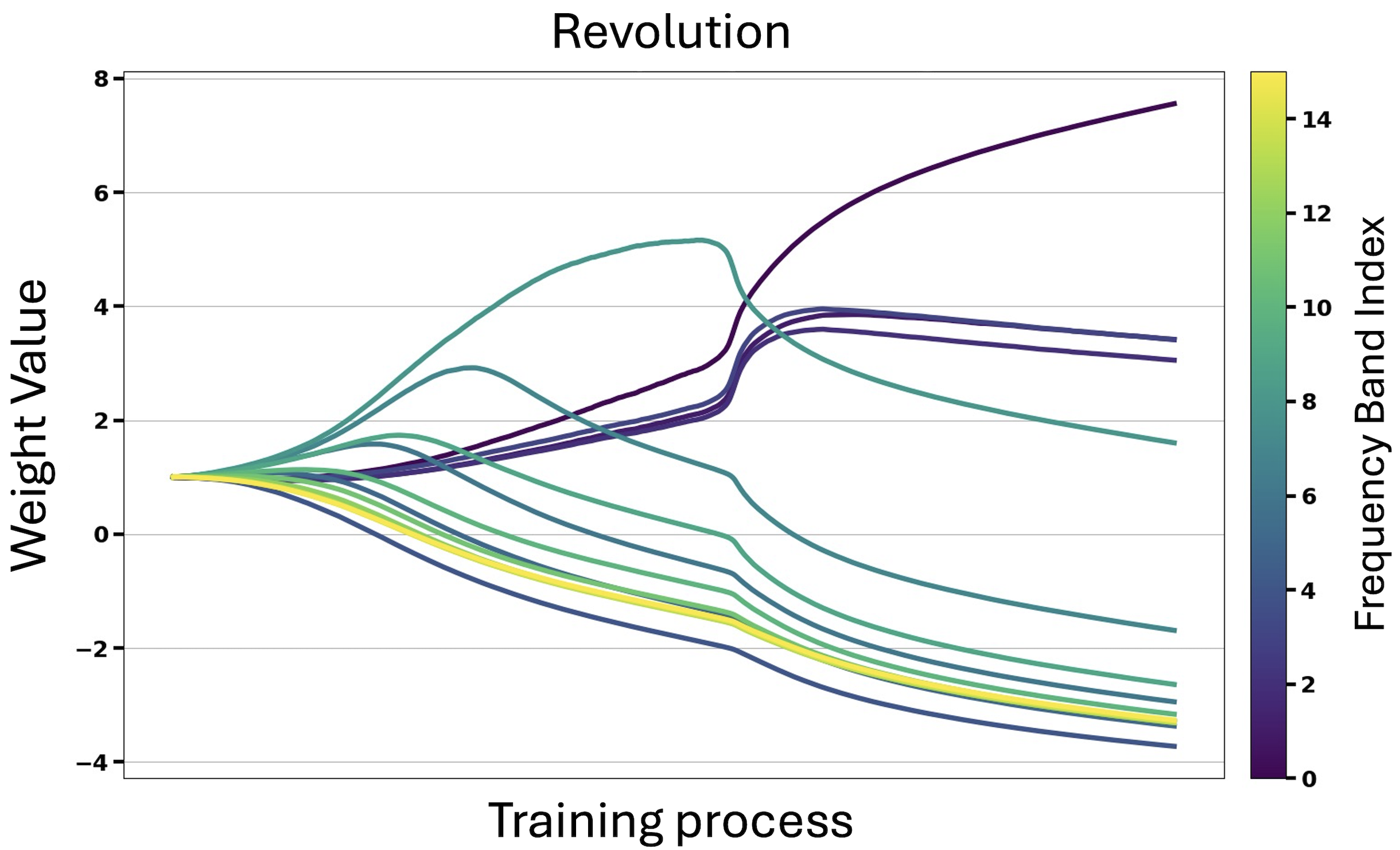}} &
\subfigure[]{\label{fig:band-weights}\includegraphics[width=0.45\textwidth]{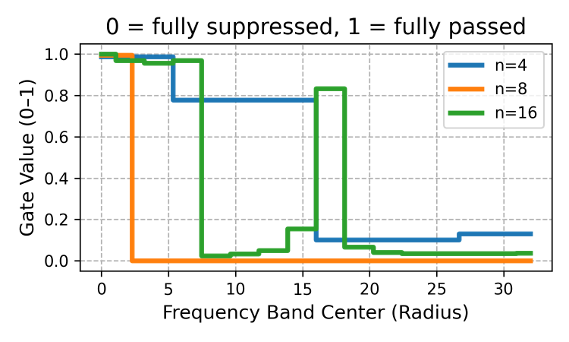}}
\end{tabular}
\caption{
(a) Training dynamics of adaptive frequency band weights for $N=16$. Each line corresponds to one of the 16 frequency bands, color-coded from low (purple) to high (yellow) frequencies. The trends show a progressive increase in low-frequency weights, transient peaks in mid-frequency bands, and gradual suppression of high-frequency components over the course of training.
(b) Final learned pass-through rates for each frequency band under FreqSelect, shown for different band counts ($N=4,8,16$). Values range from 0 to 1, indicating the relative degree of frequency preservation by each band.
}

\label{fig:train-pass}
\end{figure}

\begin{figure}[tbp]
    \centering
    \includegraphics[width=1\linewidth]{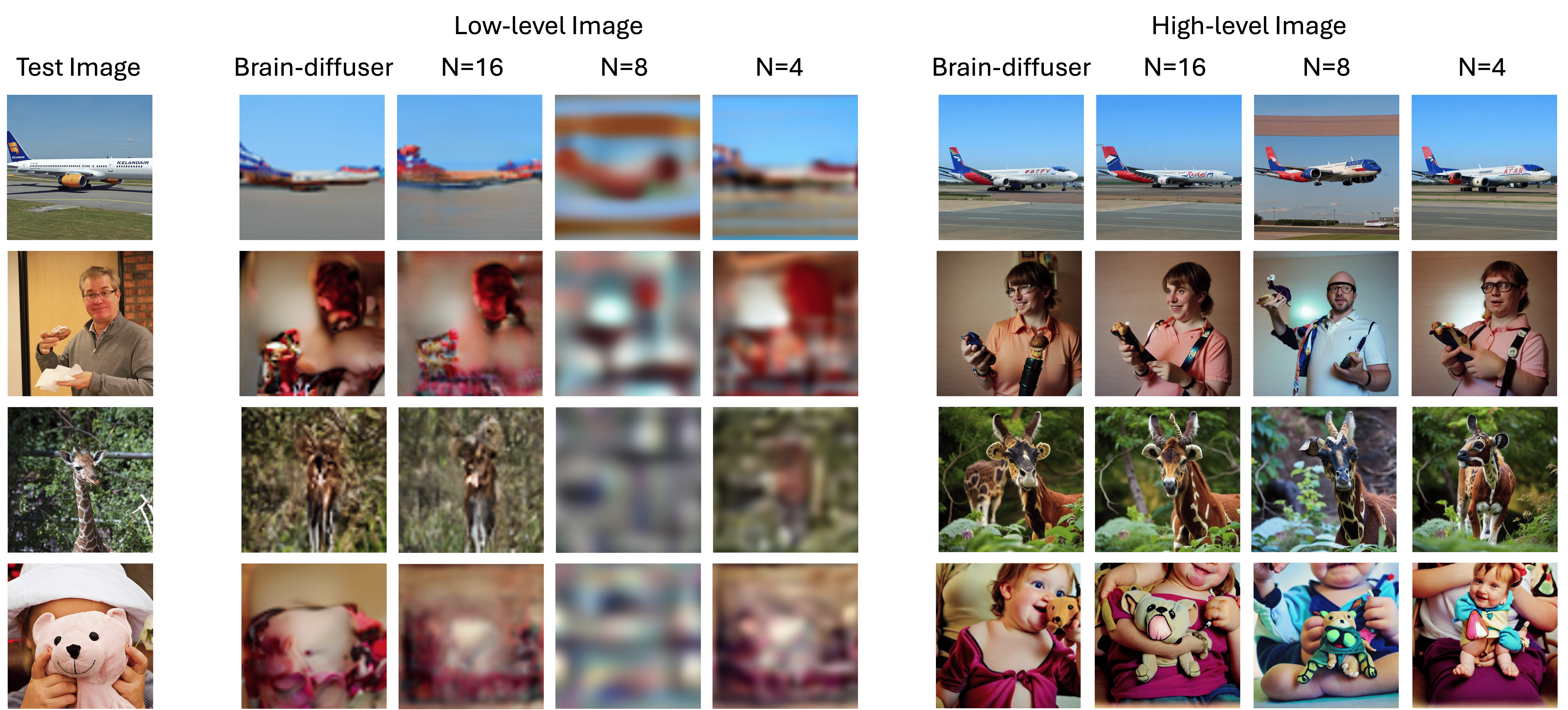}
    \caption{Comparison of fMRI reconstructions from Brain-Diffuser~\cite{ozcelik2023naturalscene} and our FreqSelect. The first column shows ground-truth images. Each panel compares Brain-Diffuser with FreqSelect using $N=4$, $8$, and $16$ frequency bands.
    }

    \label{fig:Comparison of fMRI reconstructions with Brain-Diffuser}
\end{figure}

\section{Experiment}
\label{subsubsec:Evaluation}

\subsection{Experimental Setup}
\label{app:setup}

\textbf{Dataset.} We conduct all experiments on the Natural Scenes Dataset~\cite{allen2022massive}. a 7 Tesla fMRI collection in which participants viewed images from COCO. From the original eight subjects, we selected the four (sub1, sub2, sub5, and sub7) who completed every trial for our analyses. Each image was presented for three seconds while subjects performed a continuous recognition task. The training set comprises 8,859 unique images and 24,980 fMRI trials (each image shown up to three times), and the test set comprises 982 images with 2,770 trials. we applied the General ROI mask, extracting responses from different voxels for each subjects respectively.
For our FreqSelect module we vary the number of bands $N\in\{4,8,16\}$, spacing the cutoff frequencies uniformly over $[0,32]$.  All band‐weight parameters $\vw \in \mbr{N}$ are initialized to $\mathbf{1}$, so that $\sigma(w_i)=\mathrm{sigmoid}(1)\approx0.73$ at start.

\textbf{Metric.}
In order to comprehensively examine the reconstruction performance of the model at different levels, we use the following evaluation indicators:
PixCorr quantifies pixel‐level fidelity by computing the Pearson correlation between each reconstructed image and its ground‐truth \cite{pearson1895note}, whereas SSIM evaluates perceptual similarity through comparisons of luminance, contrast, and structural components \cite{wang2004image}. Mid‐level alignment is measured via AlexNet(2) and AlexNet(5), which correlate feature activations from the 2nd and 5th convolutional layers of AlexNet, respectively \cite{krizhevsky2012imagenet}. At the semantic level, Inception Score gauges both diversity and classifiability using the final pooling outputs of Inception-v3 \cite{salimans2016improved}, while CLIP Score computes the cosine similarity between CLIP-Vision image embeddings and their corresponding text embeddings \cite{radford2021learning}. Finally, reconstruction quality in learned feature spaces is assessed by the mean squared error in EfficientNet-B's feature maps \cite{tan2019efficientnet} and by cosine similarity of SwAV-ResNet50 representations \cite{caron2020unsupervised}.



Below, we present both qualitative and quantitative results.

\subsection{Qualitative Evaluation}

\textbf{FreqSelect as a dynamic spectral gate.} We present two complementary visualizations of FreqSelect's behavior: the training-time dynamics shown in Figure~\ref{fig:Training-Time Dynamics}, and the final learned pass-through rates across different band counts in Figure~\ref{fig:band-weights}. In Figure~\ref{fig:Training-Time Dynamics}, the lowest-frequency curves (dark purple) increase steadily throughout training, indicating that the model progressively relies on coarse, low-frequency components to represent global contours and basic structure. In contrast, mid-frequency bands exhibit a temporary rise, peaking mid-training, before being de-emphasized, suggesting transient reliance on texture and edge details. The highest-frequency bands (bright green to yellow) are consistently suppressed, often reaching negative values, which aligns with our goal of filtering out fMRI-induced high-frequency noise and preventing overfitting.

Functionally, FreqSelect acts as a trainable, differentiable pre-encoder gate that modulates the spectral content entering the VDVAE–diffusion pipeline. Figure~\ref{fig:band-weights} shows the learned pass-through values (after sigmoid) for different band counts ($N=4$, $8$, $16$). Across all cases, FreqSelect assigns the highest weights to the lowest bands, reflecting the importance of global structural information in fMRI-based decoding. At $N=4$, the model relies almost entirely on the lowest band, discarding others. With $N=8$, although more bands are available, the model still suppresses most beyond the first few, indicating that the useful signal remains concentrated in low frequencies. At $N=16$, the finer spectral resolution allows a more nuanced behavior: lower bands are fully passed, middle bands receive moderate weights, and higher bands are effectively gated out.
These learned pass-through rates are bounded between 0 and 1, providing smooth, differentiable control over frequency inclusion. Critically, FreqSelect learns these rates end-to-end, adapting to spectral profile of fMRI signal without relying on static, hand-crafted filters. This flexibility allows practitioners to choose $N$ to balance frequency resolution and model complexity, ensuring that informative spectral bands are retained while noisy ones are suppressed.

\begin{figure}[tbp]
    \centering
    \includegraphics[width=\linewidth]{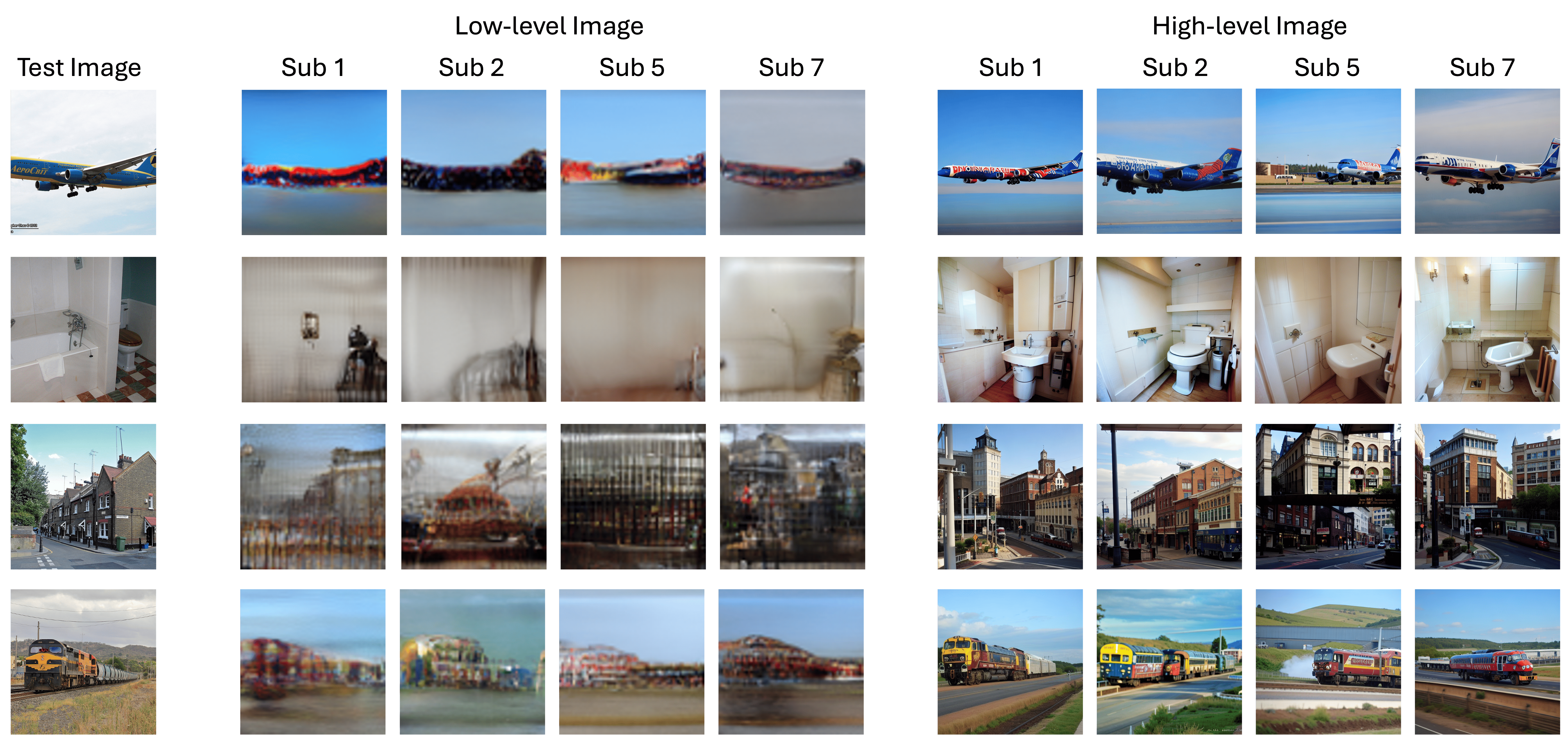}
    \caption{Example fMRI reconstructions using our FreqSelect module. The first column shows the ground-truth test image. In both the low- and high-level reconstruction panels, each subsequent column corresponds to an individual subject (Sub1, Sub2, Sub5, Sub7).
}
    \label{fig:fMRI Reconstructions between subs}
\end{figure}

\textbf{Visual and subject-level impact of band selection.} Although FreqSelect reliably preserves the lowest-frequency band regardless of $N$, increasing $N$ improves spectral resolution at the cost of model complexity. However, poor choices, especially $N=8$, can overly prune informative mid-frequency bands, degrading reconstruction quality. We analyze this trade-off in Figure~\ref{fig:Comparison of fMRI reconstructions with Brain-Diffuser}, which compares fMRI-based reconstructions across different $N$ values ($4$, $8$, $16$) and against Brain-Diffuser~\cite{ozcelik2023naturalscene}.
At $N=4$ and $N=8$, reconstructions exhibit oversmoothing, lacking texture and detail relative to Brain-Diffuser and $N=16$. In particular, the $N=8$ condition discards more mid-frequency content, yielding outputs with distorted contrast and incorrect colors. For instance, in the third row of Figure~\ref{fig:Comparison of fMRI reconstructions with Brain-Diffuser}, Brain-Diffuser reconstructs a deer image that mismatches the test image, whereas FreqSelect, benefiting from better spectral filtering, correctly captures the number of deer. In the fourth row, Brain-Diffuser erroneously inserts a face behind a teddy bear, while FreqSelect correctly omits it, demonstrating its capacity to suppress misleading high-frequency artifacts.

We further assess generalization across subjects in Figure~\ref{fig:fMRI Reconstructions between subs}, where the same test images are reconstructed for four individuals (Sub1, Sub2, Sub5, Sub7). While inter-subject variability, such as in contrast or fine detail, is evident, semantic content and spatial layout are well preserved across all reconstructions. In low-level outputs (first row of each subject panel), structural features like airplane fuselage contours, sharp bathroom fixtures, building facades, and parallel train tracks are highly consistent. This consistency suggests that FreqSelect robustly captures task-relevant spatial-frequency components, even under substantial neural variability across individuals.

\begin{wrapfigure}{r}{0.4\linewidth}
    \centering
    \centering
    \includegraphics[width=\linewidth]{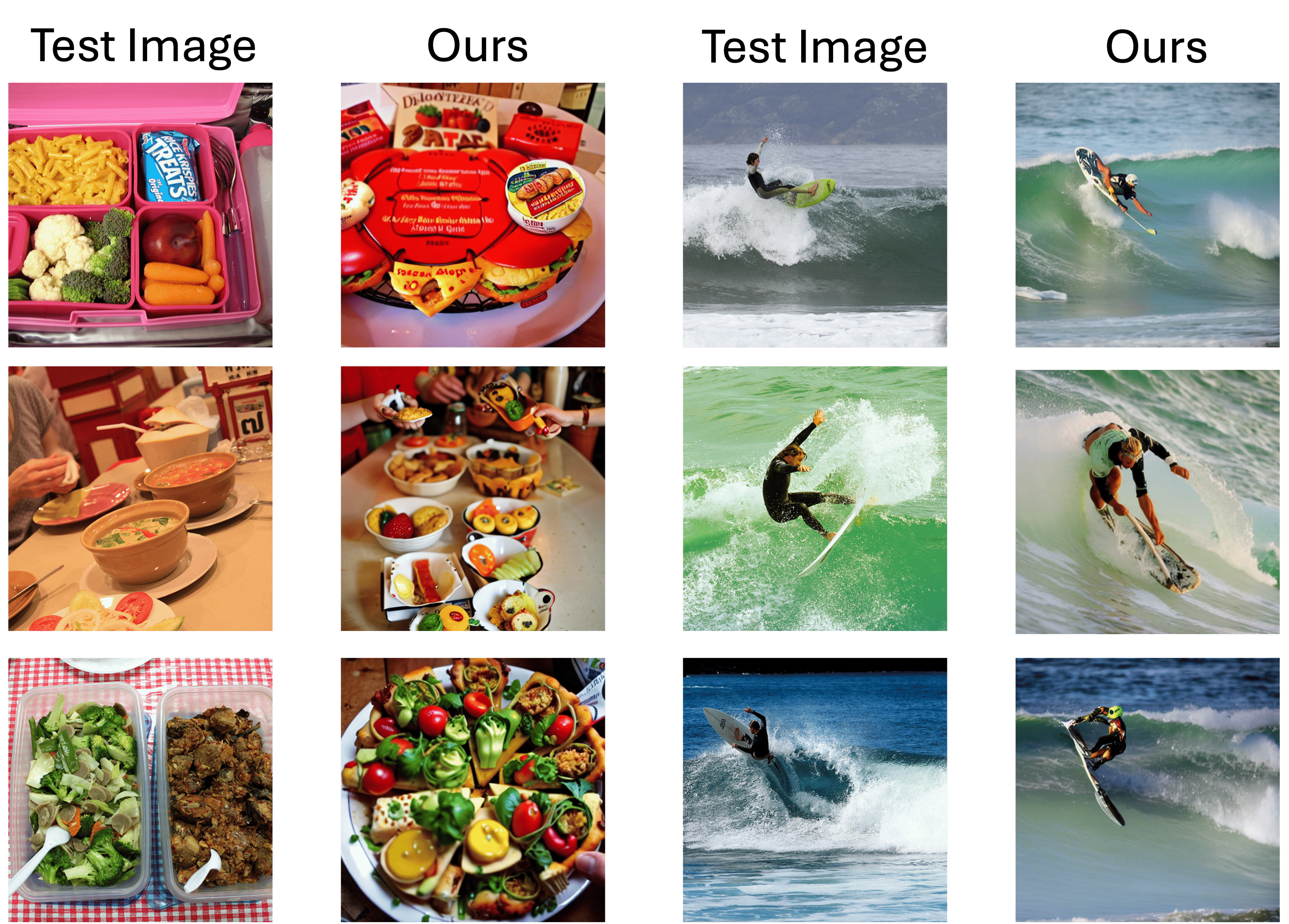}
    \caption{Failure cases from fMRI reconstructions using FreqSelect. 
    }
    \label{fig:bad_result}
\end{wrapfigure}


\textbf{Characterization of reconstruction failures.} Reconstruction errors reveal systematic challenges in capturing fine-grained semantic details despite faithful low-frequency structure recovery.
Figure~\ref{fig:bad_result} illustrates typical failure cases: regardless of the original food type, be it a colorful bento, hamburger platter, or buffet, the reconstructions degrade into a generic ``tower-shaped'' arrangement of similar snack cups and green decorations. The model loses the diverse shapes, colors, and material textures, retaining only a universal ``food combination'' template.
In contrast, for natural scenes like waves, the model successfully reproduces low-frequency features such as the texture and color gradients of seawater and the overall wave structure, demonstrating a strong capacity for capturing large-scale background patterns. However, finer details in the foreground, like human postures, facial expressions, and surfboard features, are heavily distorted or blurred. Semantic information is lost, resulting in imprecise outlines and ambiguous object identities.



\begin{table}[tbp]
\centering
\resizebox{\textwidth}{!}{%
\begin{tabular}{|l|c|c|c|c|c|c|c|c|}
\hline
\multirow{2}{*}{\textbf{Method}} 
  & \multicolumn{4}{c|}{\textbf{Low-Level}} 
  & \multicolumn{4}{c|}{\textbf{High-Level}} \\
\cline{2-9}
  & \textbf{PixCorr $\uparrow$} 
  & \textbf{SSIM $\uparrow$} 
  & \textbf{AlexNet(2) $\uparrow$} 
  & \textbf{AlexNet(5) $\uparrow$} 
  & \textbf{Inception $\uparrow$} 
  & \textbf{CLIP $\uparrow$} 
  & \textbf{EffNet-B $\downarrow$} 
  & \textbf{SwAV $\downarrow$} \\
\hline
Brain-Diffuser~\cite{ozcelik2023naturalscene} 
  & \textbf{0.304}   & 0.293    & \textbf{96.84\%} & \textbf{97.48\%} & \textbf{88.6\%}  & 92.5\%   & \textbf{0.761}  & \textbf{0.410}  \\
\hline
Ours (N=4)    
  & 0.2902  & 0.2914   & 94.94\%          & 96.80\%          & 88.11\%         & 91.85\%  & 0.7725         & 0.4176         \\
\hline
Ours (N=8)    
  & 0.0738  & 0.2633   & 86.82\%          & 93.13\%          & 86.00\%         & 91.81\%  & 0.7986         & 0.4520         \\
\hline
Ours (N=16)   
  & 0.2734  & \textbf{0.2961} & 96.25\% & 97.46\% & 88.36\% & \textbf{92.65\%} & 0.7672 & 0.4141 \\
\hline
\end{tabular}%
}
\caption{Quantitative comparison of fMRI reconstructions across different $N$ of our method and Brain-Diffuser. Best scores are bolded. Higher is better ($\uparrow$) for PixCorr, SSIM, AlexNet(2/5), Inception, and CLIP; lower is better ($\downarrow$) for EffNet-B and SwAV distances.
}
\label{tab:quantitative_comparison}
\end{table}

\subsection{Quantitative Evaluation}

\textbf{FreqSelect boosts reconstruction fidelity.} 
Table~\ref{tab:quantitative_comparison} presents a quantitative comparison between Brain-Diffuser~\cite{ozcelik2023naturalscene} and our proposed FreqSelect on subject 1, evaluated across three spectral settings ($N \in \{4, 8, 16\}$). Brain-Diffuser serves as a strong baseline, delivering high scores on both low-level metrics (PixCorr, SSIM, AlexNet) and high-level metrics (Inception, CLIP, EfficientNet-B, SwAV), as detailed in Section~\ref{subsubsec:Evaluation}.

As $N$ increases, FreqSelect gains representational flexibility. With $N=4$, it matches Brain-Diffuser in capturing coarse image structure. At $N=16$, it outperforms Brain-Diffuser on several metrics by finely suppressing high-frequency components dominated by noise, while preserving informative low-frequency content. This pattern highlights the value of fine-grained, adaptive band weighting in reconstructing images from fMRI signals.
However, when $N=8$, performance drops significantly. The model converges to retain only the lowest-frequency band, discarding mid- and high-frequency components that carry crucial structural and semantic cues. This failure stems from the coarse and uniform partitioning of the frequency domain: each band spans too wide a frequency range, mixing signal and noise. As a result, the model suppresses entire bands rather than selectively filtering noise, compromising overall reconstruction quality.

\textbf{FreqSelect achieves the best balance of detail and semantics.}
Table~\ref{tab:quantitative_comparison_withothers} presents qualitative comparisons between FreqSelect and five state-of-the-art methods. Except for Brain-Diffuser, whose reconstructions are taken from ~\cite{ozcelik2023naturalscene}, all other results are reported from their respective papers.
Lin \etal\cite{Lin_Deep_Frequency_Filtering_CVPR2023} first used the Natural Scenes Dataset for fMRI-to-image reconstruction, using a StyleGAN2 generator. Takagi \etal\cite{takagi2023high} used a latent diffusion model to produce recognizable outlines from fMRI data. Gu \etal\cite{gu2023decoding} used an instance-conditioned GAN trained on ImageNet to generate semantically coherent reconstructions. Ferrante \etal\cite{ferrante2023generative} introduced a multimodal alignment approach that co-reconstructs images and captions, yielding plausible silhouettes but lacking in low-level detail and texture.

Among all models, Brain-Diffuser~\cite{ozcelik2023naturalscene} performs consistently well across most metrics. Its two-stage architecture uses a VDVAE encoder for robust global layout and a diffusion model for photorealistic refinement.
Our FreqSelect module builds on this architecture by introducing adaptive frequency-band weighting to retain task-relevant signals and suppress irrelevant noise. This refinement addresses Brain-Diffuser's uniform frequency treatment, resulting in improved SSIM and CLIP scores. The SSIM gains stem from better preservation of structural edges and contrast, while the improved CLIP scores reflect enhanced semantic fidelity through retention of frequency content most aligned with vision–language embeddings.
%
See Appendix~\ref{app:discussion} for further discussion and Appendix~\ref{app:future} for future work.


\begin{table}[tbp]
\centering
\resizebox{\textwidth}{!}{%
\begin{tabular}{|l|c|c|c|c|c|c|c|c|}
\hline
\multirow{2}{*}{\textbf{Method}} 
  & \multicolumn{4}{c|}{\textbf{Low-Level}} 
  & \multicolumn{4}{c|}{\textbf{High-Level}} \\
\cline{2-9}
  & \textbf{PixCorr $\uparrow$} 
  & \textbf{SSIM $\uparrow$} 
  & \textbf{AlexNet(2) $\uparrow$} 
  & \textbf{AlexNet(5) $\uparrow$} 
  & \textbf{Inception $\uparrow$} 
  & \textbf{CLIP $\uparrow$} 
  & \textbf{EffNet-B $\downarrow$} 
  & \textbf{SwAV $\downarrow$} \\
\hline
Lin \etal \cite{Lin_Deep_Frequency_Filtering_CVPR2023}
  & -   & -    & - & - & 78.2\%  & -   & -  & -  \\
\hline
Takagi \etal \cite{takagi2023high}    
  & -  & -   & 83.0\%          & 83.0\%          &76.0\%         & 77.0\%  & -         & -        \\
\hline
Gu \etal \cite{gu2023decoding}    
  & 0.150  & 0.325   & -          & -         & -         & -  & 0.775        & 0.423         \\
\hline
Ferrante \etal \cite{ferrante2023generative}   
  & \textbf{0.353}  & 0.287 & 89.00\% & 97.00\% & 84.00\% & 90.00\% & - & - \\
\hline  
Ozcelikfu \etal \cite{ozcelik2023naturalscene} 
  & 0.304   & 0.293    & \textbf{96.84\%} & \textbf{97.48\%} & \textbf{88.6\%}  & 92.5\%   & \textbf{0.761}  & \textbf{0.410}  \\
\hline
\textbf{Ours} (with \textit{FreqSelect})
  & 0.2734  & \textbf{0.2961} & 96.25\% & 97.46\% & 88.36\% & \textbf{92.65\%} & 0.7672 & 0.4141 \\
\hline 
\end{tabular}%
}
\caption{Quantitative comparison of fMRI reconstructions. Best scores are in bold. 
}
\label{tab:quantitative_comparison_withothers}
\end{table}

\section{Conclusion} 

We presented \textit{FreqSelect}, an adaptive frequency-selection module that enhances fMRI-to-image reconstruction by selectively filtering spatial-frequency bands before encoding. Integrated seamlessly into VDVAE-diffusion pipelines, FreqSelect improves reconstruction quality by emphasizing frequency components most predictive of neural activity while suppressing noise. Across multiple quantitative and qualitative benchmarks on the Natural Scenes Dataset, our approach consistently outperforms existing methods in both structural fidelity and semantic alignment. Moreover, the learned frequency patterns reveal interpretable insights into how the brain processes visual information at different spatial scales. FreqSelect offers a general, lightweight framework applicable across subjects and potentially extendable to other neural recording modalities, paving the way toward more accurate and interpretable neural decoding.

\section*{Acknowledgments}
Junliang Ye conducted this research under the supervision of Lei Wang and Md Zakir Hossain as part of his final year master's research project at ANU.
This work was supported by computational resources provided by the Pawsey Supercomputing Centre, a high-performance computing facility funded by the Australian Government. We sincerely thank the anony-
mous reviewers for their invaluable insights and constructive feedback,
which have greatly contributed to improving our work.

\bibliography{egbib}

\newpage
\appendix

\section{Appendix}

\subsection{Related Work}
\label{app:relatedwork}

\textbf{Neuroscientific insights on spatial frequency representation.} Visual images can be decomposed into spatial‐frequency bands, where low frequencies capture global structure and luminance contrast, while high frequencies convey fine textures and sharp contours~\citep{devalois1982spatial}. Foundational studies using sinusoidal gratings in fMRI experiments have revealed that early visual areas exhibit selective, band‐pass responses to specific spatial frequencies~\citep{singh2000spatiotemporal,henriksson2012retinotopic}. In particular, V1 (primary visual cortex) responds most strongly to intermediate spatial frequencies that support edge detection and local detail~\citep{Mannion2015Spatial}. 
Extrastriate areas such as V2 and V3 integrate signals from V1 into larger receptive fields, which tend to broaden frequency tuning and slightly shift preference toward lower frequencies~\citep{Mannion2015Spatial}. 

While these neuroscientific findings highlight the brain's selective sensitivity to specific frequency ranges, existing neural decoding models have largely ignored this property by treating all frequencies uniformly. 
In contrast, our work is directly informed by these principles: \textit{FreqSelect} incorporates a learnable, lightweight module that dynamically emphasizes spatial-frequency bands most predictive of neural responses. By aligning model inductive biases with cortical frequency preferences, FreqSelect not only improves decoding performance but also yields interpretable maps that reflect how different spatial scales are encoded in the brain.

\textbf{Two-stage fMRI reconstruction with VAEs and diffusion models.} Early fMRI-to-image decoding work relied on linear regression and Bayesian inference to reconstruct coarse stimulus representations from voxel responses~\cite{miyawaki2008visual, naselaris2009bayesian}
. Subsequent methods introduced deep generator priors, matching DNN feature activations via GANs or DGN priors~\cite{shen2019deep}. However, these single-stage methods either have difficulty capturing high-level semantics or robustly suppressing noise. Two-stage generative models have become the dominant architecture for fMRI-based image reconstruction. These methods typically involve an initial low-resolution approximation followed by refinement with a powerful generative prior. 
For instance, the Very Deep VAE (VDVAE) introduced by Child \etal~\cite{child2021vdvae} uses a deep hierarchy of latent variables to capture multi-scale image structure, from coarse layout to fine details. Similarly, VQ-VAE~\citep{oord2017neural} uses vector quantization to discretize the latent space, which enhances feature preservation and semantic control.

Diffusion models have further improved reconstruction quality by using expressive priors in a compressed latent space. Stable Diffusion~\citep{Rombach_2022_CVPR} and Versatile Diffusion~\citep{xu2023versatile} perform denoising steps in low-dimensional latent spaces, reducing computational cost while maintaining image fidelity.
Building on these components, recent two-stage decoding pipelines, \eg, Brain-Diffuser~\citep{ozcelik2023naturalscene} and MindDiffuser~\citep{lu2023minddiffuser} have set the state of the art on the Natural Scenes Dataset. 
Brain-Diffuser maps fMRI signals to VDVAE latents to reconstruct coarse image layouts, which are then refined with Versatile Diffusion to add color, texture, and object-level details. 
MindDiffuser instead predicts VQ-VAE codes and CLIP embeddings to guide Stable Diffusion, combining structural accuracy with semantic consistency. 

However, both models uniformly process all spatial-frequency components, treating high- and low-frequency information as equally informative. This design fails to account for the fact that fMRI signals contain noise that is not evenly distributed across frequency bands. \textit{FreqSelect} addresses this shortcoming by learning to gate frequency components before encoding, enabling the model to suppress irrelevant frequencies and focus reconstruction capacity on informative ones, an approach grounded in the actual statistics of brain responses.

\textbf{Frequency filtering in vision models.}
Frequency-domain processing has gained increasing traction in vision models as a means to improve efficiency, robustness, and generalization~\citep{chi2020ffc,rahaman2019spectral}. Traditional convolutional neural networks (CNNs) operate in the spatial domain, implicitly learning filters with fixed receptive fields that impose certain frequency biases~\citep{lecun1998gradient}. Studies on spectral bias have shown that deep networks tend to prioritize low-frequency components during training, which can hinder the representation of fine details~\citep{Xu_Learning_Frequency_CVPR2020,rahaman2019spectral}. To mitigate this, some approaches use learnable Gaussian filters~\citep{Saldanha2021Frequency} or explicitly transform inputs via the Fast Fourier Transform (FFT) to retain specific frequencies while reducing spatial resolution~\citep{Xu_Learning_Frequency_CVPR2020}.

Recent work has further integrated frequency-domain operations into model architectures. 
Fast Fourier Convolution (FFC)~\citep{chi2020ffc} processes features in both spatial and Fourier domains to capture long-range dependencies, while Deep Frequency Filtering (DFF)~\citep{Lin_Deep_Frequency_Filtering_CVPR2023} applies adaptive frequency masks to enhance cross-domain generalization. 
In medical imaging, the Fourier Convolution Block (FCB)~\citep{sun2024fcb} improves MRI reconstruction by enhancing the effective receptive field using global frequency convolutions. 
Similarly, Adaptive Frequency Filters (AFF)~\citep{Huang2023AFFNet} apply learned gating over frequency-transformed token representations in vision transformers, achieving strong performance–efficiency trade-offs.

These methods, however, are designed for classification, segmentation, or supervised reconstruction tasks with full access to image labels. Crucially, none are tailored for fMRI-to-image decoding, where the input signals are noisy, indirect reflections of the original image content. 
Our proposed \textit{FreqSelect} is the first frequency-filtering module specifically optimized for neural decoding. It learns frequency gating via supervision from brain-predicted VDVAE latents, aligning its filters with neural evidence rather than image-based labels. Its modularity enables plug-and-play integration into existing pipelines without retraining large generative backbones, and the learned frequency profiles correspond to known band-pass tuning in early visual cortex, bridging computational modeling and cognitive neuroscience.

\subsection{Discussion}
\label{app:discussion}

\textbf{Frequency band insights.}
FreqSelect dynamically adapts frequency emphasis during training, enhancing reconstruction fidelity by balancing noise suppression and detail preservation.
As shown in Figure~\ref{fig:Training-Time Dynamics}, the model initially prioritizes low-frequency bands, transiently boosts mid-frequency bands to capture texture, and suppresses high-frequency noise. At the optimal number of bands $N=16$, FreqSelect achieves competitive improvements in SSIM and CLIP scores, especially in selective reconstruction scenarios (\eg, correctly omitting occluded faces unlike Brain-Diffuser~\cite{ozcelik2023naturalscene}). This demonstrates the model's ability to learn interpretable pass-through rates $\alpha_i = \sigma(w_i)$, modulating frequency bands adaptively to improve reconstruction quality.

Figure~\ref{fig:band-weights} further reveals the impact of band granularity: (i) At $N=4$, only coarse low-frequency bands survive, producing oversmoothed outputs. (ii) At $N=16$, FreqSelect finely balances mid- and low-frequency preservation against high-frequency noise, surpassing Brain-Diffuser in key metrics. (iii) At $N=8$, equal-width banding over-prunes critical mid-frequency components, degrading performance across all metrics. This underscores the importance of choosing $N$ based on the underlying fMRI power spectrum or exploring non-uniform frequency partitioning in future work.

Qualitative comparisons (Figures~\ref{fig:Comparison of fMRI reconstructions with Brain-Diffuser}, \ref{fig:fMRI Reconstructions between subs}) confirm that FreqSelect consistently produces sharper object contours, better color fidelity, and improved multi-object layouts, while generalizing robustly across subjects. Remaining failures (Figure~\ref{fig:bad_result}) highlight the limitations of overemphasizing low frequencies at the expense of high-frequency semantic details, a challenge compounded by MSE-driven training and fMRI's limited spatial resolution.

Future improvements should include auxiliary losses (\eg, perceptual, adversarial) to encourage retention of mid- and high-frequency information, adaptive or data-driven band boundaries instead of uniform splits, and extension to other modalities such as EEG~\cite{Murray2005EEG} and MEG~\cite{Baillet2017MEG}, where frequency-specific signal quality varies. These advances could further close the gap to human-level decoding fidelity and deepen our understanding of how the brain encodes visual information across spatial scales.

\subsection{Future Work}

\label{app:future}

While FreqSelect demonstrates clear benefits in fMRI-to-image reconstruction, the reviews highlight several open challenges and opportunities for future research. We outline these directions below.

\textbf{Choosing $N$ and designing better band partitions.} 
Our experiments revealed instabilities at intermediate band counts, most notably the performance dip at $N=8$, likely caused by overly coarse, uniform frequency partitions that entangle signal and noise. 
Future work will explore \emph{non-uniform, data-driven partitions} (\eg, logarithmic spacing, $k$-means on the power spectrum, or differentiable cutoffs learned end-to-end) as well as \emph{orientation-aware bands} (elliptical or steerable masks). 
To prevent the model from collapsing onto a single dominant band, we plan to incorporate \emph{anti-collapse regularizers} such as entropy penalties, Dirichlet priors, or band-dropout. We will also provide practical guidelines for selecting $N$ across data regimes, for example by relating $N$ to voxel SNR, ROI coverage, or dataset size, and reporting iso-compute frontiers to help practitioners balance accuracy and cost.


\textbf{Beyond uniform frequency gating.} 
Currently, frequency gates are global and radial. An important next step is to design \emph{spatially adaptive} gates (low-resolution gate maps predicted from the image or neural input) and \emph{multi-scale variants} that operate at different encoder resolutions. Such designs may preserve critical mid-frequency content (textures and edges) without re-introducing high-frequency noise, improving the model's ability to capture structural detail.

\textbf{Stronger and broader baselines.} Our present implementation uses a frozen VDVAE encoder for comparability. Future work will examine unfreezing early VDVAE layers or inserting parameter-efficient adapters (\eg, LoRA, bottleneck adapters) to allow end-to-end fine-tuning with FreqSelect. 
We will also evaluate alternative generative backbones such as VQ-VAE/VQ-GAN or MAE, and perform ablations on the number of latent layers used. This will test the robustness of FreqSelect across different encoders and strengthen the claim of generality.

To contextualize FreqSelect, we will compare against (i) fixed, hand-crafted band-pass filters, (ii) FFT-attention or adaptive frequency convolution networks, and (iii) no-gate baselines matched for compute. These experiments will clarify the unique advantages of adaptive, brain-informed gating over existing frequency-aware modules.


\textbf{Objective functions for mid-/high-frequency retention.} 
To address cases where fine semantic details collapse despite good low-frequency reconstructions, we will incorporate auxiliary objectives such as perceptual feature losses, frequency-domain consistency losses, or lightweight adversarial losses. These objectives explicitly encourage retention of mid- and high-frequency information while maintaining low-frequency denoising.

\textbf{Generalization across datasets, modalities, and subjects.} 
Thus far, FreqSelect has been validated only on NSD. We will extend evaluation to other naturalistic fMRI datasets, more diverse stimulus sets, and even additional neural modalities such as EEG and MEG, where spectral SNR differs from fMRI. We also plan to examine cross-subject transfer, for example via subject-agnostic gates combined with subject-specific adapters, and analyze ROI-wise band preferences to link interpretability with known visual neuroscience findings.

\textbf{Toward real-time feasibility.} 
Although real-time decoding was not a focus of this work, we will explore low-latency variants of FreqSelect by precomputing masks, streaming FFTs, and implementing early-exit inference in the diffusion stage guided by fMRI-predicted CLIP features. We will quantify the accuracy-latency trade-off to assess online applicability.


Collectively, these directions address how to (i) design better partitions, (ii) balance spectral resolution with computational cost, (iii) generalize across models, datasets, and modalities, and (iv) mitigate mid-band suppression and failure cases. Pursuing these avenues will ensure that FreqSelect remains both practical for neural decoding applications and valuable for neuroscientific insight.

\end{document}